\documentclass[aps,pre,twocolumn]{revtex4}
\usepackage{amsmath,bm}\usepackage{epsfig}
\usepackage{color}%\usepackage{graphicx}

\begin{document}
\sloppy

\newlength{\plotwidth}
\setlength{\plotwidth}{0.48\textwidth}

\title{Turbulence appearance and non-appearance in thin fluid layers}

 \author{Gregory Falkovich$^{1,2}$ and Natalia Vladimirova$^3$}

\affiliation{$^{1}$Weizmann Institute of Science,
Rehovot 76100 Israel\\$^{2}$Institute for Information Transmission Problems,
Moscow, Russia\\$^3$University of New Mexico, Albuquerque, USA}

\date{\today}
\begin{abstract}
Flows in fluid layers are ubiquitous in industry, geophysics and
astrophysics. Large-scale flows in thin layers can be considered
two-dimensional (2d) with bottom friction added. Here we find that the
properties of such flows depend dramatically on the way they are
driven.  We argue that wall-driven (Couette) flow cannot sustain
turbulence at however small viscosity and friction.  Direct numerical
simulations (DNS) up to the Reynolds number $Re=10^6$ confirm that all
perturbations die in a plane Couette flow. On the contrary, for
sufficiently small viscosity and friction, we show that finite
perturbations  destroy the pressure-driven laminar (Poiseuille)
flow. What appears instead is a traveling wave in the form of a jet
slithering between wall vortices. For $10^4<Re<5\cdot10^4$, the mean
flow has remarkably simple structure: the jet is sinusoidal with a
parabolic velocity profile, vorticity is constant inside vortices,
while the fluctuations are small. At higher $Re$ strong fluctuations
appear, yet the mean traveling wave survives. Considering the momentum
flux barrier in such a flow, we derive a new scaling law for the
$Re$-dependence of the friction factor and confirm it by DNS.
\end{abstract}

%\pacs{47.27.-i, 47.10.+g, 47.27.Gs}

 \maketitle
%\section{Introduction}
Century and a half of ever-expanding studies of turbulence onset in
three-dimensional (3d) channel and pipe flows brought a wealth of
fundamental and practical knowledge, see e.g. \cite{Rev} and the
references therein. The wall-driven flow is always
linearly stable, while the bulk-force-driven flow can be linearly unstable
for Reynolds numbers large enough  \cite{Lin,OK}. Notwithstanding this
difference and irrespective of linear stability, all flows undergo transition to turbulence at
sufficiently high Reynolds numbers when finite-amplitude
perturbations persist
\cite{Rev}. In a pipe flow, some perturbations can take a form of
traveling waves of finite amplitude \cite{FE,WK,BH1}, yet all
patterns are unstable and transient, so that the 3d flow is quite
irregular already at moderate Reynolds numbers \cite{HH,Rev}.

In contrast, for quasi-two-dimensional channel flows %in fluid layers,
it is not even known if they are able to produce turbulence at
all. This is despite a rapidly expanding interest in such flows
motivated by the needs of industry, astrophysics, geophysics, and
laboratory experiments in fluid layers and soap films (see e.g. \cite{soap1,soap2}, the
recent collection \cite{PhysFluids} and the numerous references therein).
To the best of our knowledge, in all experiments
in  layers and films, external forces and obstacles of
elaborate geometry were needed to produce turbulence (see e.g. \cite{Pinaki}), and it is not
known if such turbulence is able to sustain itself in a channel
flow past an obstacle. The reason is that 2d ideal hydrodynamics
%possesses an infinity of local conservation laws, in particular,
conserves two quadratic
invariants, energy (squared velocity) and enstrophy (squared
vorticity). Force  at intermediate scales can generate
two-cascade turbulence with energy/enstrophy cascading respectively
upscales/downscales. On the contrary, the input is at the largest
scale in a wall or pressure-driven channel flow, so that it is apriori unclear what kind of
turbulence, if any, can exist in the limit of low viscosity and friction.

\begin{figure}
\begin{center}
\includegraphics[width=\plotwidth]{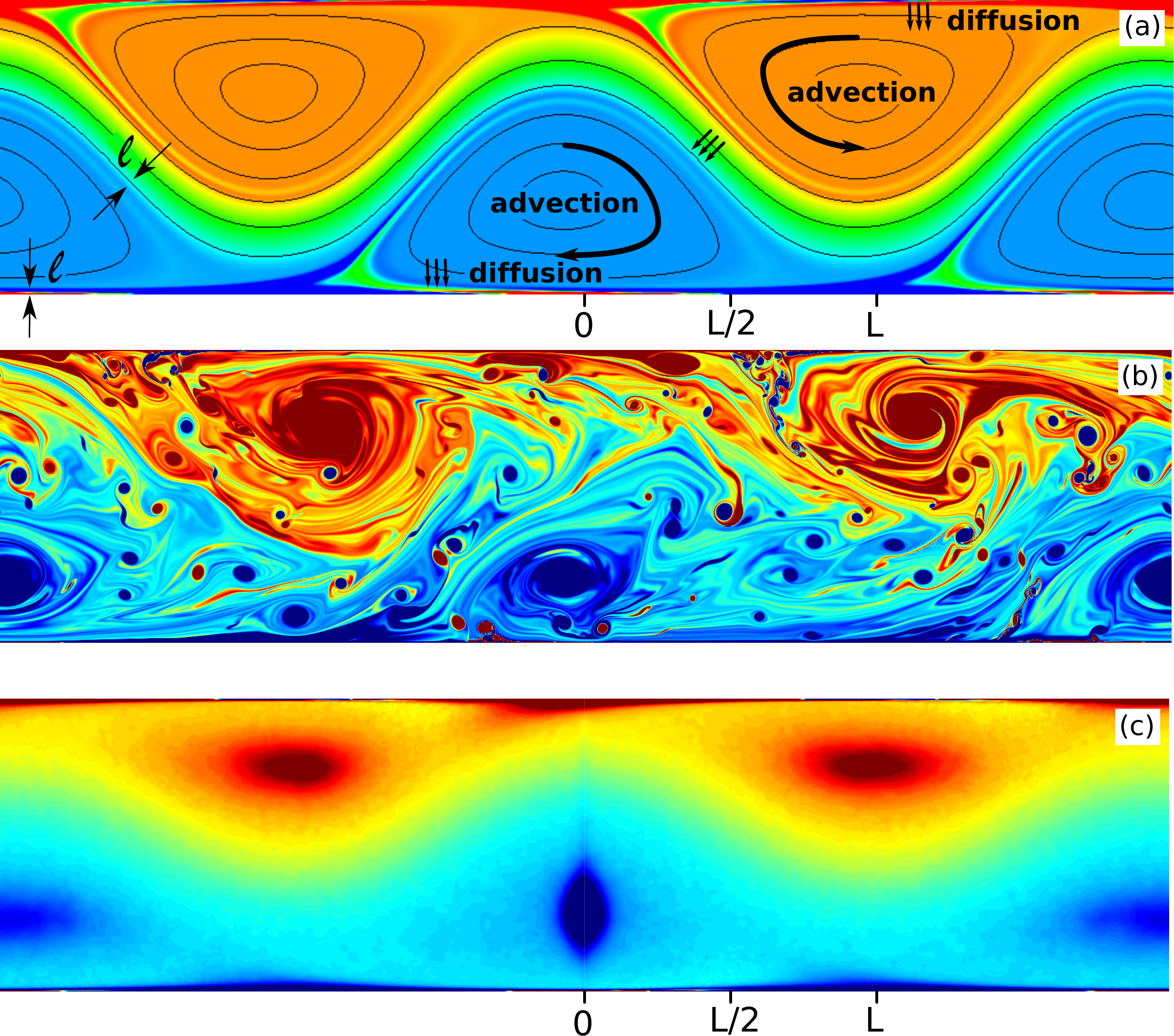}
\end{center}
\caption{ Pressure-driven flows at moderate and large Reynolds number. Vorticity snapshot for
 %in the reference frame of the traveling wave
 (a)   $Re = 1.89\cdot 10^4$
 with streamlines,
 (b)  $Re  = 3.92 \cdot 10^5$;
 (c) vorticity averaged over 7000 snapshots  over propagation distance $1400 L$ in the frame of the
 negative vortex for  $Re = 3.92 \cdot 10^5$.  }
\label{images}
\end{figure}

In this work, combining analytic theory and DNS, we answer this
fundamental question. We find out when and how turbulence appears in
pressure-driven flows: as ``snake'' --- a traveling wave in the form of a jet
meandering between counter-rotating vortices and preserving its form even
for strong fluctuations, as shown in Figure~\ref{images}. Even more
remarkably, we find that wall-driven flows remain laminar
forever. Both findings substantially widen our fundamental perspective
on turbulence and may lead to diverse practical applications.

Turbulence appearance and non-appearance in two-dimensional
hydrodynamics probably can be best illuminated by considering
interplay between momentum and vorticity, bearing in mind that
vorticity is the velocity derivative across  a
uni-directional flow. Convection carries vorticity unchanged while
viscosity diffuses it, so that any turbulence must lead to vorticity
mixing and homogenization. We thus expect the mean vorticity profile
in a turbulent flow (outside viscous boundary layers at the walls) to
be more flat than the laminar profile. On the other hand, turbulence
transfers momentum better than a laminar flow, thus increasing
drag and decreasing velocity in the bulk. These two requirements
are in a perfect agreement for pressure-driven flows
where turbulence flattens both the velocity and vorticity mean
profiles.  On the contrary, the mean velocity profile is monotonous
for wall-driven flows, so decreasing velocity in the bulk while
keeping it at the walls would make the vorticity profile more
non-uniform. We then conclude that momentum and vorticity requirements
on turbulence in 2d wall-driven flows are contradictory.

Large-scale flows in thin layers can be effectively described as
two-dimensional with bottom friction added. Consider 2d Navier-Stokes
equation with unit density and the uniform friction rate $\alpha$:
\begin{equation} \partial_t{\bm v} +({\bm v}\cdot\nabla){\bm
	v} = \nu\Delta{\bm v}-\nabla p-\alpha {\bm v},\quad
\nabla \cdot {\bm v} = 0.
\label{Navier-Stokes-alpha}
\end{equation}
Already for the simplest frictionless case (relevant e.g. for flows on
superhydrophobic surfaces \cite{Rot} or soap films under low air pressure)
dramatic difference from the three-dimensional case
is apparent.  Denote $u,v$ the fluctuating velocity components respectively parallel
and perpendicular to the mean flow $U(y)$ directed parallel to the walls, which are
placed at $y=\pm L/2$ and move with $\pm V/2$. Temporal average of (\ref{Navier-Stokes-alpha})
with $\alpha=0$ can be written using vorticity: $\omega=\nabla\times{\bm v}$
 and $\Omega=-\partial_yU$
  \begin{eqnarray}
    && \partial_y(\nu\Omega+\langle
uv\rangle)=\nu \Omega_y  -\langle v\omega\rangle=
     - \partial_x  \langle p\rangle =A.
    \label{Planar1}
  \end{eqnarray}
Turbulence adds extra  fluxes of horizontal momentum and vorticity, related by the
Taylor theorem: $\partial_y\langle uv\rangle=-\langle
v\omega\rangle$. We see that with zero pressure gradient $A$, presence of
turbulence would absurdly mean that the vorticity flows against the
mean vorticity gradient. One can also argue that the laminar profile $U=Vy/L$
already has a constant vorticity; one cannot excite turbulence
to make it more flat.
Adding to the viscous flow extra dissipation due to bottom friction
could only diminish fluctuations but cannot create them.

\begin{figure}
\begin{center}
  \includegraphics[width=\plotwidth]{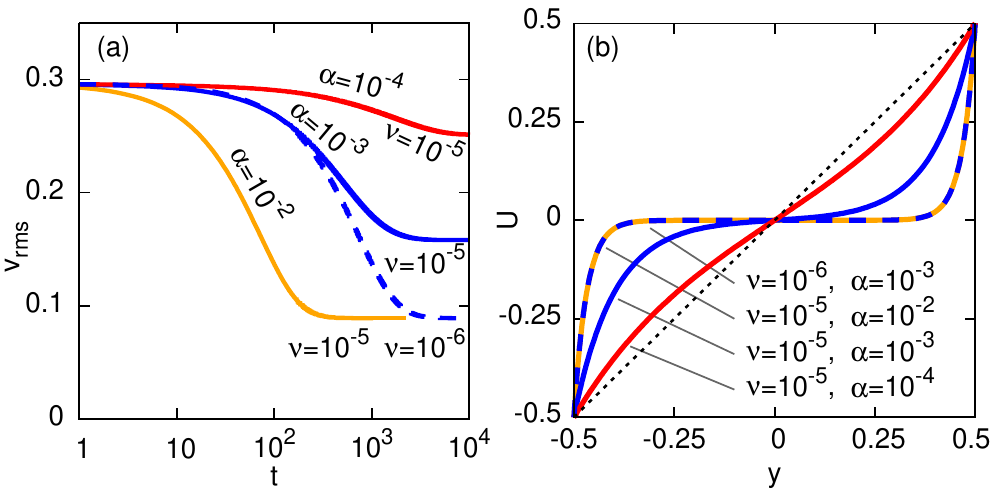}
\end{center}
\caption{ Relaxation of wall-driven flows with $V=1,L=1$ and different
  viscosities and friction to the laminar profile: (a) RMS velocities
  as function of time; (b) profiles of saturated flow,
indistinguishable from the laminar solution.
  %Here, $L=1$, $u |_{y=\pm 0.5} =\pm 0.5$.
  Initial perturbation had 8 vortex layers and amplitude
  of 0.1.}
\label{Couette}
\end{figure}

These non-rigorous but plausible arguments suggest that a wall-driven
flow must relax to the laminar state,
$U(y)=V\sinh(y\sqrt{\alpha/\nu})/2\sinh(L\sqrt{\alpha/\nu}/2)$, for
any $\nu$ and $\alpha$. That conclusion is supported by DNS whose details
are described in the Supplementary Information
(SI).  Starting from different multi-vortex configurations, we
observe different transients and eventual relaxation to the laminar flow.
Higher friction makes relaxation more monotonous and
controls the relaxation time, while the final energy is determined by
$\alpha L^2/\nu$, as follows from the laminar solution, see Figure~\ref{Couette}.
%As expected, the
%laminar solutions approach linear profile as $\alpha L^2/\nu$ decreases.
Apparently, the laminar wall-driven flow  is the global attractor in two dimensions.
To the best of our knowledge, this is the first such example in the whole fluid mechanics.

In thin 3d layers, an ability of moving walls to excite turbulence
must depend on the layer thickness $h$. We
expect turbulence when the wall Reynolds number $Vh/\nu $ becomes
large. How the wall-generated 3d turbulence will be distributed over a
wide channel deserves future studies, particularly on account of the
tendency of strong planar flows to suppress vertical motions
\cite{Shats}. Note that $\alpha=3\nu/h^2$ for planar laminar flows
with open surface and no-slip bottom, while vertical motions makes the
very notion of $\alpha$ unapplicable.

From another perspective, impossibility of turbulence in a 2d wall-driven flow
can be  related to a sign-definite mean vorticity  and a mean shear.
Even if one initially creates (as we did)
vortices of both signs, the opposite-sign vorticity is getting
destroyed by the shear while the same-sign vorticity is getting
homogenized back into the laminar profile.  On the contrary, for the
pressure-driven flow, the mean vorticity has opposite signs
at opposite walls, so that turbulence cannot homogenize vorticity
back to the laminar profile.

We turn now to the pressure-driven flow and define the respective
dimensionless control parameters $Re_A=A^{1/2}L^{3/2}/\nu$ and
$Ru_A=A^{1/2}/\alpha L^{1/2}$. Here $A$ has a dimensionality of
acceleration (force per unit mass), it is either pressure gradient
divided by density or gravity acceleration for soap films.

There is a rich history of modeling 2d Navier-Stokes channel flows with the goal to
find threshold $Re$ over which finite perturbations can be sustained,
see, e.g. \cite{OK,Jimenez} and the references therein.  To the best
of our knowledge, the largest $Re=10^4$ was achieved in
\cite{Jimenez}, where transitional turbulence was observed and it was
estimated that fully developed turbulence is to appear around
$2\cdot10^5$ which were beyond computer resources back then. Here we
explore higher $Re$ never treated before; we also add uniform friction
to relate to applications in real fluid layers.

We observe that the transition to the steady-state is slow and can be
non-monotonic, some transient behavior is discussed in SI.  In all
cases we find that pressure-driven flows relax to either of two
states: the laminar uni-directional flow or a traveling wave with much
slower average flow rate.  In the latter case, most of the flux occurs
along a sinusoidal jet meandering between two sets of counter-rotating
vortices rolling along the walls, Fig.~\ref{images}. While we
cannot rule out that both laminar and sinuous states are
long-living meta-stable states (as in 3d pipe flow),
we have not seen switches between %excursions from
them once the steady state is established.

\begin{figure}
\begin{center}
  \includegraphics[width=\plotwidth]{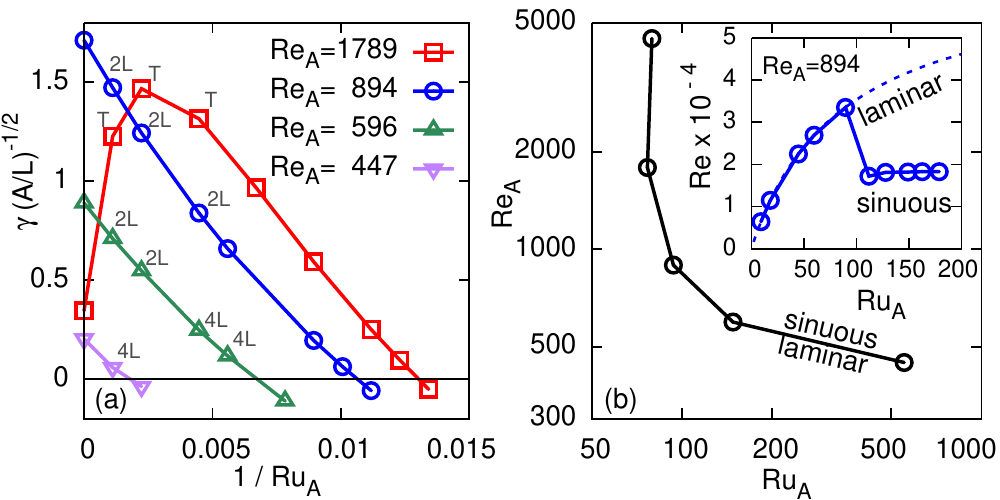}
\end{center}
\caption{Stability of the pressure-driven flow. (a) The %dimensionless
  growth rate of perturbations with $\lambda = 4L$ versus dimensionless
  friction for different viscosities.  Marks 4L and 2L indicate wavelengths
  of the long-term sinuous flow, ``T'' indicates long-term turbulent flow.  (b)
  Stability diagram in the friction-viscosity space.  Inset: Reynolds
  number versus friction. }
\label{re-ru}
\end{figure}

The time of transients can be reduced by starting with a
low-amplitude, large-scale perturbation to the laminar profile to
mimic naturally developing instability (presumably, the one predicted
by Lin for the case of $\alpha=0$ \cite{Lin}).  Then, the early
evolution shows well-defined exponential growth, suggesting linear
instability. Modeling a $12L$ section of the channel, we
applied perturbations with the wavelengths $\lambda_{\rm pert} = 3, 4,
6, 12L$ for $Re_A = 894$ and $Ru_A = 179$. The largest growth rate
$\gamma$ was found for $\lambda_{\rm pert} = 4L$, so we used this
wavelength for all other simulations; the results are shown in
Fig.~\ref{re-ru}a for different $Re_A$ and $Ru_A$.  The line in the
plane $Re_A-Ru_A$ with zero $\gamma$ separates laminar and sinuous
flows in Fig.~\ref{re-ru}b. Note that friction stabilizes the laminar
flow.  The inset in Fig.~\ref{re-ru} shows the Reynolds number,
$Re=\overline{U}L/\nu$, based on the mean flow rate, $\overline{U}$,
as a function of $Ru_A$ for $Re_A = 894$. When friction is large
($Ru_A$ is small), the flow is laminar.
%with flow rate given by Eq.~\ref{poi_lam_avg}.
As friction is reduced, the laminar flow becomes faster, yet, at $Ru_A
\approx 100$ the system transitions to the sinuous state and the flow
rate drops. That means that one can speed up the flow by increasing
friction, which facilitates transition from the sinuous to the laminar
regime.

The long-time evolution of the instability shows that right after the threshold   the sinuous
flow appears on the scale of the domain, $\lambda = 4L$, while at higher $Re_A,Ru_A$ we
observe transition to a sinuous flow with wavelength $\lambda =2L$.
%, independently of the wavelength of the initial perturbation and the length of the channel, $4L$ or $12L$.
These long-term states are indicated in Fig.~\ref{re-ru}a. Reducing dissipation even
further, we observe an unsteady, sinuous-like flow with strong
fluctuations, which we call ``turbulent state''. Below, we
take a closer look at these long-term states state for frictionless systems.

\begin{figure}
  \includegraphics[width=\plotwidth]{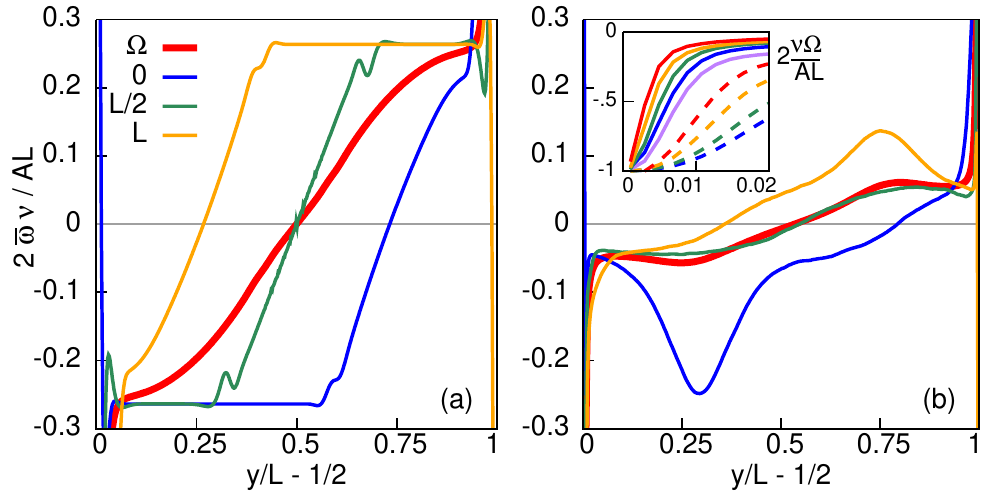}
\caption{Vorticity $\overline\omega(x,y)$ time-averaged in the moving
  frame of a vortex.  Vertical slices are taken through $x=0$, $L/2$, and $L$, as shown in
  Fig.~\ref{images}, for $Re_A=516$ (a) and $Re_A = 4000$ (b). The thick red line is zonally averaged
  vorticity $\Omega(y)=\int\overline\omega(x,y)\,dx$.  Inset: $\Omega$ in the boundary layer
  for $Re_A= 516$, 362, 894, 1265, 2000, 2828, 4000, 5656, and 8000,
  from bottom to top. }
\label{slices}
\end{figure}

Without friction,
we observe uni-directional Poiseuille flow up to $Re_A \approx 500$
and $Re \approx 2\cdot 10^4$. At $Re_A\approx500$ the snake is born already with a
finite amplitude of modulation (see SI).  For $500 \lesssim Re_A \lesssim 1500$,
the flow is a sinuous traveling wave with $10^4 \lesssim Re \lesssim
5\cdot 10^4$ and no noticeable fluctuations.  The flow has a
beautifully simple stationary structure in a co-moving reference
frame: The jet is sinusoidal with a parabolic velocity profile, while
vorticity is essentially constant across each vortex, which appears as
a plateau in the vorticity cross-sections  in Fig.~\ref{slices}a.
%Constant vorticity means that the stream function is quadratic in the coordinates plus harmonic function that describes potential flow.
%Quadratic streamfunction would correspond to elliptical orbits, which is not the case, as seen
%from  Figure~\ref{images}, that is the flow has substantial potential component.
Constant vorticity inside the vortices can be explained in the spirit of \cite{Bat} as a
consequence of viscosity being very small and yet finite: The former means that vorticity must be constant {\it along} the
(closed) streamlines, while the latter mean that vorticity must be
constant {\it across} the streamlines in a stationary
flow. The same argument suggests that
the flux of vorticity must be constant across the jet, and
thus vorticity must change linearly between opposite values at the
separatrices.

At $Re_A \approx 1500$ the flow becomes turbulent.  For $Re_A \gtrsim
2000$ and up to 8000 ($Re = 3.92 \cdot 10^5$) the relative level of
velocity and vorticity fluctuations remains constant within the
accuracy of our measurements. All turbulent flows have a pronounced
large-scale structure of a jet and $2L$-periodic chain of vortices,
similar to the sinuous flow. This is seen from comparison of
Fig.~\ref{images}a with Fig.~\ref{images}c, where the averaging is done
in the frame of the stronger negative vortex (the other three vortices are
blurred to a different degree by fluctuations).  In the turbulent state,
the chaotic small-scale vortices are created at the walls, pulled
out, and eventually each merges into a big vortex of the same sign
thus feeding the large-scale flow.  While horizontally averaged
velocity and vorticity for sinuous and turbulent are of similar shape
(see SI for more detail), the time-averaged flows expose qualitative
difference: in the turbulent state, mean vorticity peaks at the
centers of vortices rather than being flat across the vortex, see Fig.~\ref{slices} and Fig.~\ref{images}.

\begin{figure}
\begin{center}
  \includegraphics[width=\plotwidth]{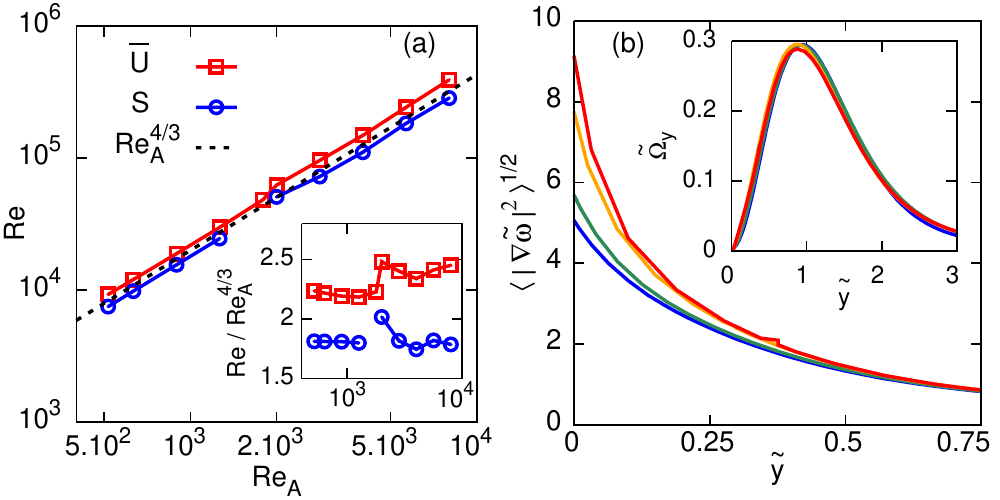}
\end{center}
\caption{ (a) The Reynolds numbers, based respectively on the %RMS velocity  $v_{rms}$,
  flow rate $\overline{U}$ and the speed of the
  traveling wave $S$, versus the Reynolds number based on the applied pressure.  Inset:
  data compensated by $Re^{4/3}$. (b) Fluctuations of the vorticity gradient
  in the boundary layer for $Re_A = 8000$, 5656, 4000, and 2828,
  from top to bottom, in coordinates $\nabla\tilde{\omega}
  =\nabla \omega Re_A^{-2/3} \nu/AL$ and $\tilde{y} = (y/L - 1/2) Re_A^{2/3}$.
  Inset: gradient of zonally averaged vorticity in the same coordinates
  for the same $Re_A$. %(There is a potential confusion here: not clear
%  if the derivative is taken with respect to $y$ or $\tilde{y}$.)
}
\label{scaling}
\end{figure}

From the topology of the mean flow profile, depicted in
Fig.~\ref{images}a, we now derive the relation between the applied
gradient $A$ and the mean flow rate $\overline{U}$ in the limit of
large $Re_A=A^{1/2}L^{3/2}/\nu$. For that, one needs to describe the
transfer of momentum (or equivalently vorticity) from the center to
the walls, taking into account two separatrices, one separating the 
vortex from the jet and another from the wall boundary layer. Vorticity is
diffused by viscosity across the separatrix, then is carried fast by
advection inside the vortex, and then transferred by viscosity towards
the wall. There are thus two viscous bottlenecks (transport barriers) in this transfer: on the jet-vortex
separatrix and on the wall boundary layer. The width $\ell$ of the separatrix boundary layer
can be estimated requiring the diffusion time
$\ell^2/\nu$ to be comparable to the turnover time $L/U$, which gives
$\ell\simeq (\nu L/U)^{1/2}$ and the effective viscosity (momentum
diffusivity) $\nu_e\simeq U\ell\simeq \sqrt{\nu UL}$ (we do not
  distinguish $U$ and $\overline{U}$ in the estimates). Requiring
that the momentum flux due to pressure gradient is carried by the
viscosity towards the walls, $A\simeq \nu_e U/L^2$, we obtain the
expressions for the effective Reynolds number and the effective
turbulent viscosity:
\begin{equation}
    Re={UL\over\nu}\simeq {L^2A^{2/3}\over\nu^{4/3}}=Re_A^{4/3}\,,\quad
   \nu_e\simeq \nu Re_A^{2/3}\ .\label{nue}
\end{equation}
To describe the wall boundary layer, note that $v\equiv0$ at a wall.
Therefore, integrating (\ref{Planar1}) over $y$ from wall to wall, we
obtain $\Omega(L/2)=-U'(L/2)=AL/2\nu$, that is always equal to the
laminar value, see the inset in Fig.~\ref{slices}b. Now we estimate
the width of the wall boundary layer, $U/U'(L)\simeq
LA^{2/3}\nu^{-1/3}/AL\nu^{-1}\simeq \nu^{2/3}/A^{1/3}\simeq \ell$,
which confirms that our estimate (\ref{nue}) is self-consistent.

Appearance of the thin boundary layer at large $Re$ (see SI for the details) must
lead to a sharp maximum of the vorticity derivative, which we estimate
as follows: $\max\Omega_y\simeq\Omega(L/2)/\ell\simeq
LA^{4/3}\nu^{-5/3}$. That value is much larger than
$\Omega_y(L/2)=A/\nu$, derived from taking (\ref{Planar1}) at a
wall. Away from the wall boundary layer, turbulence must suppress the
mean vorticity gradient, as indeed seen in the insets in
Figs~\ref{slices}b and \ref{scaling}b.

In the presence of friction the scaling
law (\ref{nue}) is expected to hold when the typical time
of the momentum transfer to the wall, $U/A\simeq L(A\nu)^{-1/3}$, is
shorter than the friction time $\alpha^{-1}$, otherwise the uniform
friction dominates and we have linear regime with $U\propto A$. This
means that the pressure acceleration must exceed both viscous and
friction thresholds: $A\gg \nu^2L^{-3}, (\alpha L)^3/\nu$.

Numerical simulations support (\ref{nue}), see Figure~\ref{scaling}.
The scaling $Re \propto Re_A^{4/3}$ continues through both regimes, of
weakly and strongly fluctuating flows, even though the proportionality
constant might be slightly changing at the transition, as seen in the
inset in Figure~\ref{scaling}a. The mean vorticity profile at the boundary layer
also follows the scaling (\ref{nue}), as shown in the inset in
Figure~\ref{scaling}b plotted for the rescaled quantity
$\tilde{\Omega}_y= \Omega_y(y)/\max\Omega_y=\nu\Omega_y/ARe_A^{2/3}$.

Let us briefly discuss the role of the turbulent fluctuations. Flow dissipates
energy and enstrophy, and the viscous dissipation rate of the former is
proportional to the latter: $\nu\langle |\nabla v|^2\rangle=\nu\langle
\omega^2\rangle$. It follows from  (\ref{nue}) that the work $A\overline{U}$ done by
the pressure gradient gives the same estimate as the energy dissipation rate
by the mean flow: $\nu \overline{\Omega^2}\simeq \nu U^2/\ell L\simeq
A^{5/3}L\nu^{-1/3}\simeq AU$.
This means that the mean flow is able to dissipate energy by itself.
%We see that the mean flow must be able to dissipate energy.
Indeed, the DNS data show (see SI, Fig. 2) that the turbulent enstrophy
fluctuations are smaller than the enstrophy of the mean flow, while
velocity fluctuations are negligible.
 The dissipation of the second inviscid invariant, enstrophy,  is determined by
 the vorticity gradients shown in Figure~\ref{scaling}b.
%We see that  the contribution of near-wall fluctuations into the vorticity
% gradients is much larger and grows with $Re$ faster than the vorticity
% of the mean flow, which follows (\ref{nue}), as shown in the inset
% plotted for the rescaled quantity $\tilde{\Omega}_y=
% \Omega_y(y)/\max\Omega_y=\nu\Omega_y/ARe_A^{2/3}$.  This suggests that
% turbulence dissipates enstrophy rather than energy, which deserves
% future research with much better statistics and resolution.
%Dissipation of vorticity fluctuations and dissipation of mean
%vorticity in the boundary layer are shown in Fig.~\ref{scaling}b and
%in the inset respectively.
We observe that the mean vorticity gradient follows
(\ref{nue}), while the contribution of near-wall fluctuations into the
vorticity gradients, and therefore into the enstrophy dissipation, is much
larger and grows with $Re$ faster than the vorticity of the mean flow.
This suggests that enstrophy is dissipated by turbulence rather than by the mean flow,
which deserves future research with much better statistics and
resolution.

One can recast (\ref{nue}) as the statement that the friction factor
of a 2d ``pipe'',  $AL/U^2$, decays as $Re_A^{-2/3}\sim Re^{-1/2}$, which is
faster than in three dimensions, where one finds the empirical Blasius
law $Re^{-1/4}$ for moderate $Re$ and the logarithmic decay for large
$Re$. Within purely 2d system (where separatrix is responsible for the
transport barrier), the scaling law (\ref{nue}) may be expected to hold at
arbitrary large $Re$, yet in fluid layers its validity is restricted
by the requirement that $\ell\simeq LRe_A^{-2/3}=LRe^{-1/2}$ exceeds
the fluid depth $h$.  Therefore, we expect that as $Re$ approaches
$(L/h)^{2}$, the decay of the friction factor with $Re$ slows down and
eventually converges to the 3d values observed in rectangular ducts
\cite{FR}.

Traveling wave pattern thus  enhances effective viscosity and suppresses the
flow rate compared to the laminar regime. It is instructive to compare the
viscosity enhancement (\ref{nue}) with the enhancement of diffusivity $\kappa$
by the factors $Pe^{1/2}$ for cellular flow \cite{Childress, BS} and
%enhancement of diffusivity along the channel by the factor
$Pe^{1/3}$ for wall-attached flow \cite{2/3}, where $Pe=UL/\kappa$.
% characterizes the ratio of diffusion and advection times the same way $Re$ does this for
%viscosity and advection.
That enhancement of diffusion leads to
acceleration of flame fronts \cite{NV} and other phenomena.
%Recall also that the first example of the enhancement in diffusion due to the
%interplay with regular advection by a laminar flow is known as Taylor
%diffusion, which concerns longitudinal spreading of passive scalar
%{\it along} the pipe, so that the effective diffusivity scales as an
%inverse of the molecular diffusivity \cite{Taylor}.
Similar to  (\ref{nue}), %our $1/3$-scaling of the effective viscosity with molecular viscosity,
interplay between small noise and advection universally leads to the
$1/3$-scaling with noise amplitude: for tumbling frequency of a
polymer in a shear flow \cite{KoT}, for the Lyapunov exponent of an
integrable system under stochastic perturbation \cite{Kurchan}.

To conclude, we have established that wall-driven 2d flow is always
laminar. We described the traveling wave pattern which replaces the
laminar flow for pressure-driven flows.  In distinction from 3d, the
traveling wave we find in 2d is stable; as the Reynolds number grows,
the fluctuations increase yet the mean flow preserves its
traveling-wave ``snake'' form. Another remarkable property of 2d snake
is that it contains separatrices, which modify momentum transport to
the walls and thus lead to a new type of scaling law for the friction
factor.

We thank A.~Obabko for help in using {\tt Nek5000}.
The work was supported by the grants of Israel Science Foundation and
the Minerva Foundation
% with funding from Federal German Ministry for Education and Research
and by NSF grant no. DMS-1412140. Simulations
are performed at Texas Advanced Computing Center (TACC) using Extreme
Science and Engineering Discovery Environment (XSEDE), supported by
NSF Grant No.  ACI-1548562 through allocation TG-DMS140028.

%\section{Supplementary Information}
\section*{SUPPLEMENTARY INFORMATION}

\subsection*{Numerical setup}
\label{numerics}

We solve the Navier-Stokes equation with bottom friction, Eq.~(1) from the main text, using a spectral element
code, {\tt Nek5000} in the domain $-L/2<y<L/2$, $0<x<L_x$ with $L=1$
and periodic boundary conditions in $x$-direction.  Unless specified,
$L_x = 4L$.  For wall-driven flows, the velocity at the boundaries is
$u(x,\pm L/2,t)=\pm u_w/2$ with $u_w=1$.  For pressure-driven flows,
no-slip boundary conditions at the walls are used.

Initial velocity has
the form ${\bm v}(x,y,0) = U_0(y){\bf i}_x + {\bm
  v}_0(x,y)$, with perturbation,
\begin{eqnarray}
%   u = - 2a \sin(\pi n y) \cos(\pi n y) \sin(\pi n x), \qquad
%   v =    a \sin(\pi n y) \sin(\pi n y) \cos(\pi n x).
   u_0 &=& - 2(a/k_x) \sin(k_y y) \cos(k_y y) \sin(k_x x), \\
   v_0 &=&  \quad \; (a/k_y)  \sin(k_y y) \sin(k_y y) \cos(k_x x),
\label{ic}
\end{eqnarray}
where $k_x=\pi n/L$, $k_y=\pi m/L$ and $n,m=1,2,3...$.
Note that the perturbation vanishes at the boundaries, satisfies
$\nabla \cdot {\bm v}_0 = 0$, and has the structure of vortex layers.

Unless specified, for wall-driven flows we use the mesh of $256\times80$
elements with 7 collocation points per element in each direction. For
pressure-driven flows we use either the same mesh with 7 or 9 collocation
points or the mesh of $512\times160$ elements with 7 collocation
points. The size of elements varies in $y$-direction to provide denser
mesh near boundaries.  The number of collocation points per element in
one direction corresponds to the spatial interpolation order.  The
method is 3rd order accurate in time.

\subsection*{Simulations of wall-driven flows}
\label{wall-driven}

For the wall-driven flows, the velocity perturbation has been added to
the linear velocity profile $U_0 = u_w y/L$. We use vortices with
aspect ratio one, $k_y=k_x=\pi n$, where $n$ corresponds
to the number of vortex layers in $y$-direction.  We have considered
viscosity down to $\nu = 10^{-6}$, perturbation amplitudes
$a/k_x = a/k_y = 0.1$, $0.5$, and $1.0$, and 4, 8, or 16 vortex layers.

There are interesting transients in the evolution of finite
perturbations in the wall-driven flow. At short times, $t \sim 1$,
merging of vortices could result in the short-term increase of maximum
velocity. One can explain it by linear model: the modulus of the sum
of the (non-orthogonal) eigen modes can increase with time even when
modes are decaying with different rates. At intermediate times, the
flow appears chaotic, yet we clearly observe the upscale energy
transfer as the number of vortices reduces with time while their size
increases.  Some vorticity is generated at the walls and propagates
inwards, but this effect weakens with time.  In the long term, all
perturbations die and the flow approach the laminar solution $U(y) =
\frac{u_0}{2} \sinh\left[y\sqrt{\frac{\alpha}{\nu}}\right] /
\sinh\left[\frac{L}{2}\sqrt{\frac{\alpha}{\nu}}\right]$ of the steady
equation $\nu U'' - \alpha U = 0$.

\subsection*{Simulations of pressure-driven flows}
\label{pressure-driven}

\begin{figure*}
  \includegraphics[width=1.75\plotwidth]{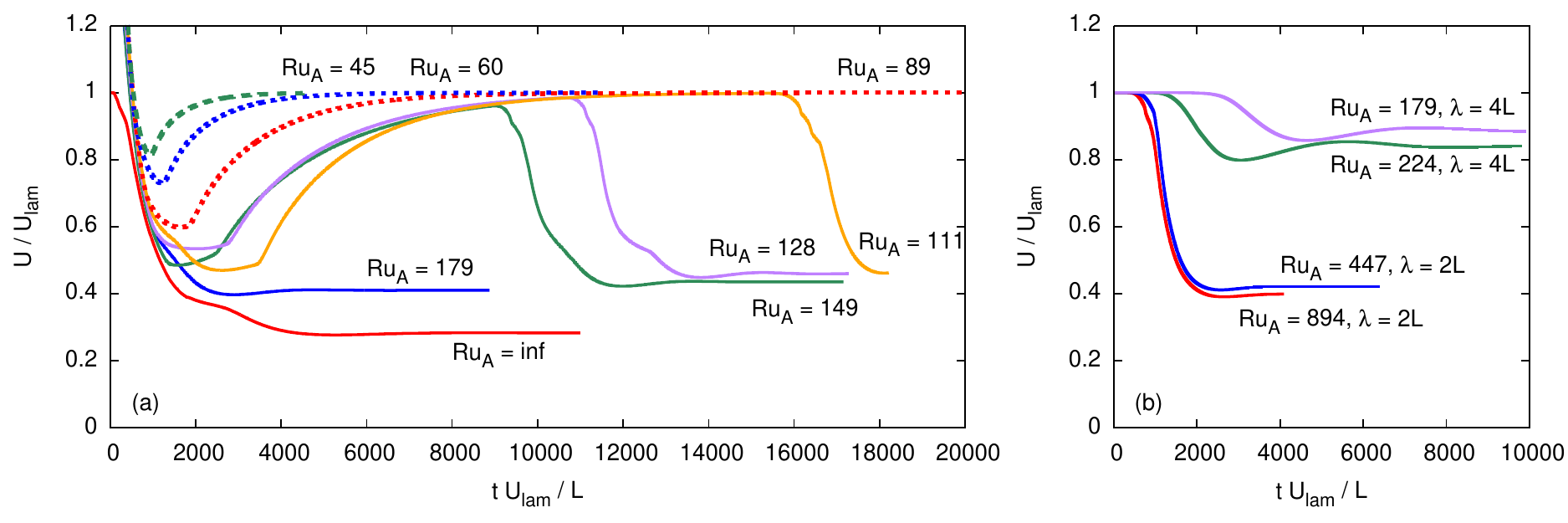}
  \caption{Flow rate $\overline{U}$ in pressure-driven systems with friction for (a)
    small-scale perturbation on top of parabolic profie for $Re_A =
    894$, and (b) large-scale perturbation on top of laminar profile
    for $Re_A = 596$.  For all cases in (a) laminar sinuous flows had
    wavelength $\lambda = 2L$; in (b) both $\lambda = 2L$ and
    $\lambda = 4L$ were observed. Simulations that evolved to sinuous
    and laminar flows are shown with solid and dashed lines
    respectively.}
    \label{evol}
\end{figure*}

We started with a numerical setup similar to the one for wall-driven
flow, except that we have enabled external pressure gradient and set
wall velocities to zero.  The initial perturbations with $k_y=k_x=\pi
n$, $n=8$, and $a/k_x = a/k_y = 0.1$ or 0.01 were added to $U_0 = 1 - 4y^2$
in the domain with aspect ration $4:1$.  We considered $Re_A=896$
($L=1$, $\nu=10^{-5}$, $A=8\times 10^{-5}$) and $Ru_A > 45$ ($0 \le
\alpha \le 2\times 10^4$).

The transient behavior took frustratingly long times and often defied expectations.
For instance, in the run for $Ru_A= 128$ shown
in Fig.~\ref{evol}a, we observed a convincing sinuous state reached at
$t \approx 1000 L/U_{\rm lam}$, where $U_{\rm lam}$ is the laminar
flow rate, yet at $t \approx 3000 L/U_{\rm lam}$ it started to
accelerate toward the laminar state, and it did so until $t \approx
11000 L/U_{\rm lam}$, when it started to decelerate. At $t \approx
16000 L/U_{\rm lam}$ the system was back to the sinuous state.  The
evolution in such complex cases is sensitive to the initial
conditions. The initial vortices, which have the size of a fraction of
the channel, do not evolve directly to the traveling wave solution.
These perturbations die, and then the remaining shear flow develops an
instability.  It occurs for sufficiently high $Re_A$ via creation of
system-size vortex pair with the wavelength of the length of box,
$\lambda_{\rm inst}=4 L$, which later transitions to the sinuous
solution with shorter wavelength, $\lambda_{\rm wave}=2L$.  The
transition to the system-size vortex pair can be detected in vertical
velocity, as low as 12-14 orders of magnitude below the horizontal
component. That means that such systems in reality may be very sensitive to the level of noise and vibrations.

We utilize this mechanism of naturally-developing instability to study
stability of the laminar flow with friction.  Instead of small-scale
vortices, we initially impose a large-scale, small-amplitude
perturbation that mimics the early stages of transition from laminar to
sinuous flow. In particular, we use a single layer of elongated
vortices, $k_y=\pi$, $k_x=2\pi n /L_x$, with the typical amplitude of
perturbation $a/\pi= 10^{-4} \, u_{0,\rm lam}$, where $u_{0,\rm lam}$ is
the velocity at the center of laminar profile. In most
simulations, we started with one pair of initial vortices in the box
with $L_x= 4L$; a small number of simulations was done with
$n=1,2,3,4$ in the box of $L_x = 12L$ (and proportionally extended
computational grid).  The background velocity was given by the laminar
profile,
\begin{equation}
u_{\rm lam} (y) = \frac{A}{\alpha} \left( 1-
  \frac{\cosh \sqrt\frac{\alpha}{\nu} y} {\cosh \sqrt\frac{\alpha}{\nu} \frac{L}{2} }
  \right),
\label{poi_lam}
\end{equation}
with the following, based on flow rate, Reynolds number,
 \begin{equation}
  Re = Re_A Ru_A\left(1 - \sqrt\frac{4Ru_A}{Re_A}\tanh\sqrt\frac{Re_A}{4Ru_A} \right).
\label{poi_lam_avg}
 \end{equation}
% \begin{equation}
% U = \frac{A}{\alpha} \left( 1 - \frac{1}{p} \tanh p \right)
% = \frac{AL^2}{4\nu} \left( \frac{1}{p^2} - \frac{1}{p^3} \tanh p \right),
% \qquad
% p\equiv \sqrt{\frac{\alpha}{\nu}} \frac{L}{2}.
%\label{poi_lam_avg}
% \end{equation}
% When $p \ll 1$, then $U\approx AL^2/12\nu$.
% When $p \gg 1$, then $U\approx A/\alpha$.
% In terms of $Re=\sqrt{AL^3}/\nu$, the dependents is
% \begin{equation}
%  \frac{U}{\sqrt{AL}} = \frac{c}{4}\left(1 - \sqrt\frac{c}{Re}\tanh\sqrt\frac{Re}{c} \right),
%  \qquad \text{where} \qquad
% c=\frac{4}{\alpha}\sqrt\frac{A}{L}.
% \label{lam_alpha}
% \end{equation}

\begin{figure}
  \includegraphics[width=\plotwidth]{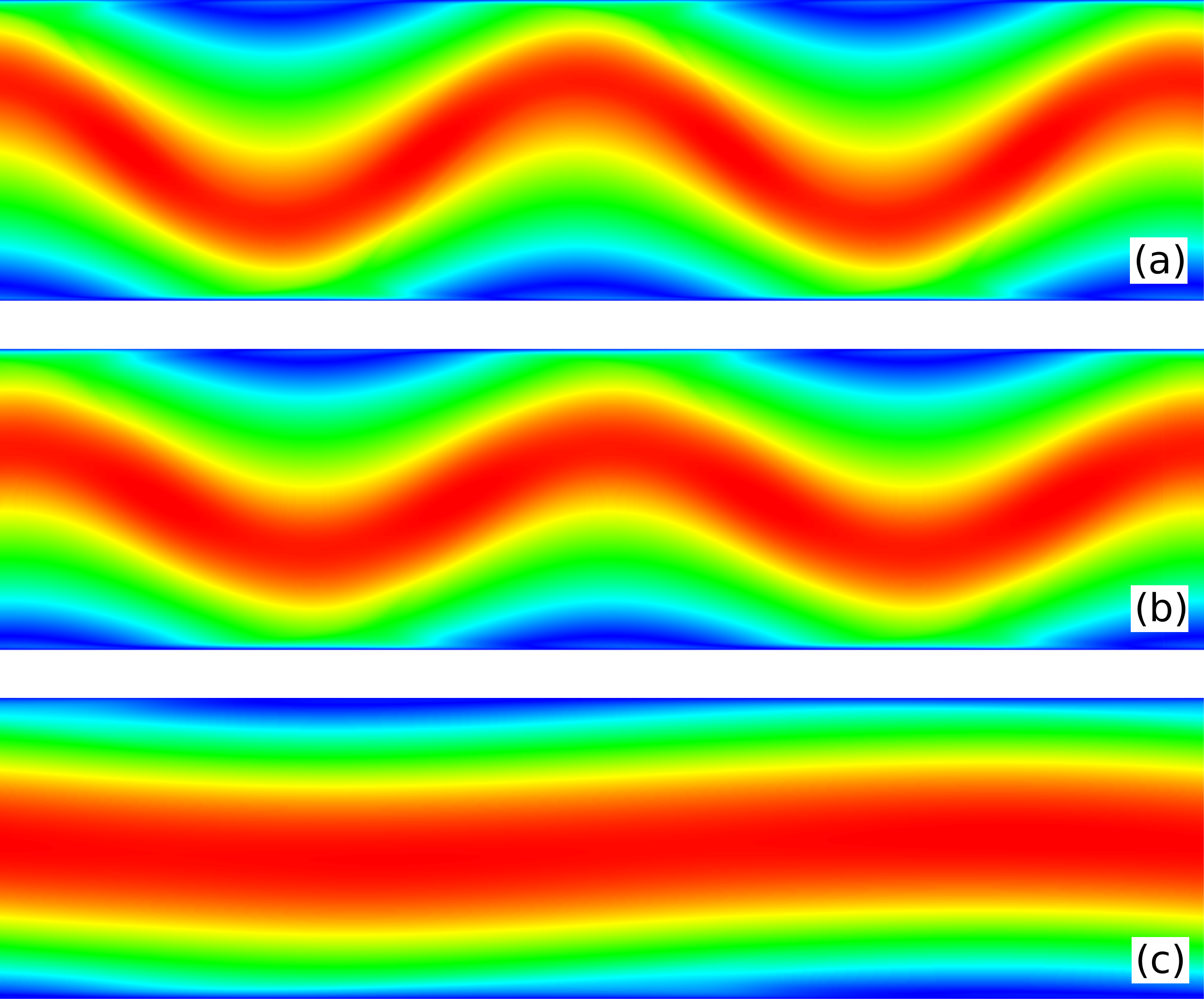}
  \caption{Snapshots of velocity modulus for
   (a) $Re_A = 894$, $\alpha = 0$ for established traveling wave,
   (b) $Re_A = 596$, $Ru_A = 179$ taken at 70\% of the time taken in Fig.~2a,
   and (c) $Re_A = 596$, $Ru_A = 447$ taken at 50\% of the time taken in Fig.~2a.
   In each case, the colormap is scaled between zero and maximum velocity.
  }
    \label{vmag}
\end{figure}

Selection of the initial velocity as the laminar flow plus small
perturbation produced more regular evolution. Early stages show
exponential growth or decay, measured by RMS of vertical velocity. As
the perturbation grows, it is transformed into the sinuous flow with
the same wavelength $\lambda = 4L$ when there is a significant bottom
friction, as seen in Figure~\ref{vmag}c.  At lower bottom friction,
the saturated state appears with doubled number of vortices, $\lambda
=2L$, seen in Figure~\ref{vmag}a,b. It may well be that in an open
channel, unrestricted by periodic boundary conditions, the
wavelengthes
% for the maximal growth rate and for the saturated state
will be different and depend on
 $ Re_A,Ru_A$.
In the main text, we present the growth rate for $\lambda=4L$ for
$Re_A = 447$, 596, 894, and 1789, (at $A=8\times 10^{-5}$) and
$Ru_A < 81$, and the stability diagram in $Re_A$-$Ru_A$ space.

\begin{table}
\begin{tabular}{rrrrrrrrrrrrr }
\hline
  $Re_A$&
  $\quad  \nu \quad$ &
  $\quad   A  \quad$ &
  $\quad   \overline{U} \quad$ &
  $\quad   S        \quad$ &
  elements &
  \ p      &
  $\;t\overline{U}/L$ &
  n
  \\  % $u_{rms}$   $\omega_{\rm rms}$  run name
\hline
 516  &  3e-5  &  2.4e-4  &  0.278  &  0.225  & 256x64  & 8 & 5.1 &  200 \\ % pgrad1/pg3e5_n8_a01b
 632  &  2e-5  &  1.6e-4  &  0.241  &  0.197  & 256x64  & 8 & 4.8 &  200 \\ % pgrad1/pg2e5_n8_a01
 894  &  1e-5  &  8.0e-5  &  0.189  &  0.156  & 256x64  & 8 & 3.8 &  200 \\ % pgrad1/pg1e5_n8_a01
1264  &  5e-6  &  4.0e-5  &  0.149  &  0.123  & 256x64  & 8 & 3.0 &  200 \\ % pgrad1/pg5e6_n8_a01b
1788  &  5e-6  &  8.0e-5  &  0.242  &         & 256x64  & 8              \\ % pgrad1/pg5e6_n8_a01a
2000  &  2e-6  &  1.6e-5  &  0.125  &  0.102  & 256x80  & 9 &  1800 & 9000 \\ %  0.147  &  0.997 &  pg2e6\\
2828  &  1e-6  &  0.8e-5  &  0.096  &  0.073  & 256x80  & 9 &  1800 & 9000 \\ % 0.108  &  0.838 &  pg1e6a\\
4000  &  1e-6  &  1.6e-5  &  0.146  &  0.111  & 256x80  & 9 &  1800 & 9000 \\ % 0.167  &  1.479 &  pg1e6b\\
5656  &  1e-6  &  3.2e-5  &  0.243  &  0.183  & 512x160 & 7 &  1400 & 7000 \\ %  0.276  &  2.655 &  pg1e6c\\
8000  &  5e-7  &  1.6e-5  &  0.196  &  0.143  & 512x160 & 7 &  1400 & 7000 \\ % 0.218  &  2.280 &  pg5e7\\
\\
\end{tabular}
\caption{Parameters of simulations of pressure-driven flows without
  friction together with obtained flow rate $\overline{U}$ and the speed of the
  traveling wave $S$.  Here, $p$ in the number of collocaton points per spectral
  element in each direction, $t$ is the time inverval used for
  averaging, and $n$ is the number of snapshots. In all cases, $L_y =
  L = 1$ and $L_x = 4L$.  In simulations with $Re_A=5656$ and 8000,
  1\% damping of hyper viscosity type was applied to the highest
  mode.}
\end{table}

To study the established flows at highest possible Reynolds numbers, we extended selected
simulations with $\alpha=0$ to times much longer than a typical time of
fluctuations in total energy and enstrophy. Then, we used the final velocity snapshots
 as initial conditions for runs at higher $Re_A$.
Summary of simulations is shown in the Table.  The table includes
simulation parameters, $A$ and $\nu$, information on numerical grids,
the resulting flow rate $\overline{U}$ and the speed of the traveling wave $S$,
and the time interval and the number of
snapshots used for averaging. As discussed in the main text, fluctuations
in velocity become noticeable only for $Re_A \gtrsim 1500$; these
simulations required averaging over long effective propagation
distances, $\sim\!10^3L$, to collect meaningful statistics.  In
contrast, to estimate average quantities and the level of
fluctuations in saturated sinuous flows with low fluctuations it was
sufficient to collect statistics for only several propagation
wavelengths.

\begin{figure}[b!]
  \includegraphics[width=\plotwidth]{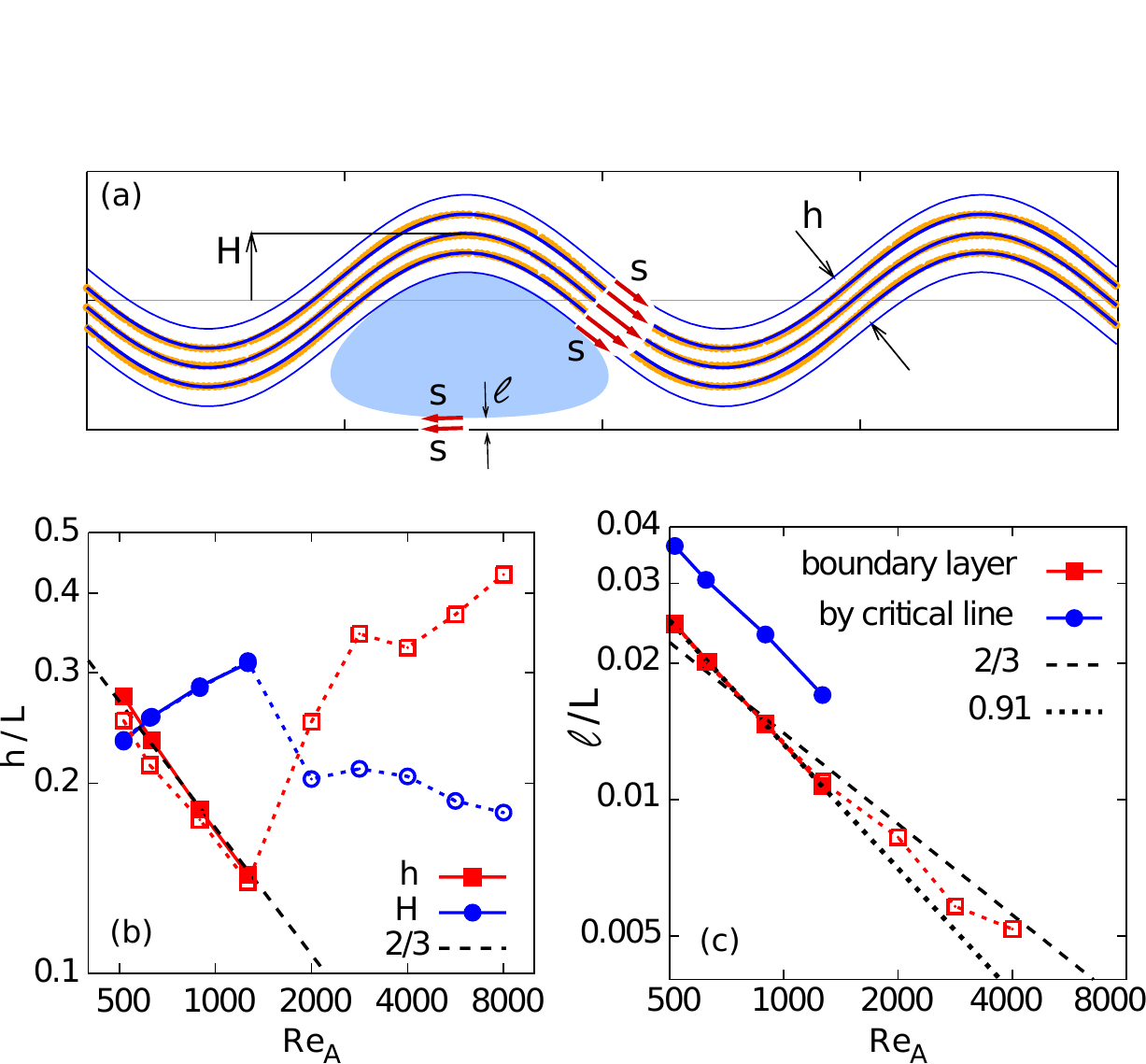}
\caption{ (a)Vorticity level sets $\omega = 0$, $\pm \omega_0/2$
  (orange dots) for $Re_A = 632$ fitted by $y = H \sin(\pi x/L) +
  const$ (blue lines).  The blue region schematically represents the
  area of constant vorticity,  the red arrows represent the flow.
  (b) Pressure dependence of the amplitude and the width of the jet.
 (c) The width of the boundary layer, $\ell$ and the minimal distance
  of the critical line from the wall.  Filled symbols and
  solid lines are from instant snapshots of the sinuous flow.
  Empty symbols and dashed lines are from
  time-averaged snapshots of the turbulent flow, where the level sets
  $\omega=0$ and $\omega_2/2$ were used,  $\omega_2$ being the maximum
  vorticity at the vertical slice through the center of positive
  vortex excluding the boundary layers, $-3/8 < y/L < 3/8$.   }
\label{jet}
\end{figure}

\begin{figure*}
  \includegraphics[width=2\plotwidth]{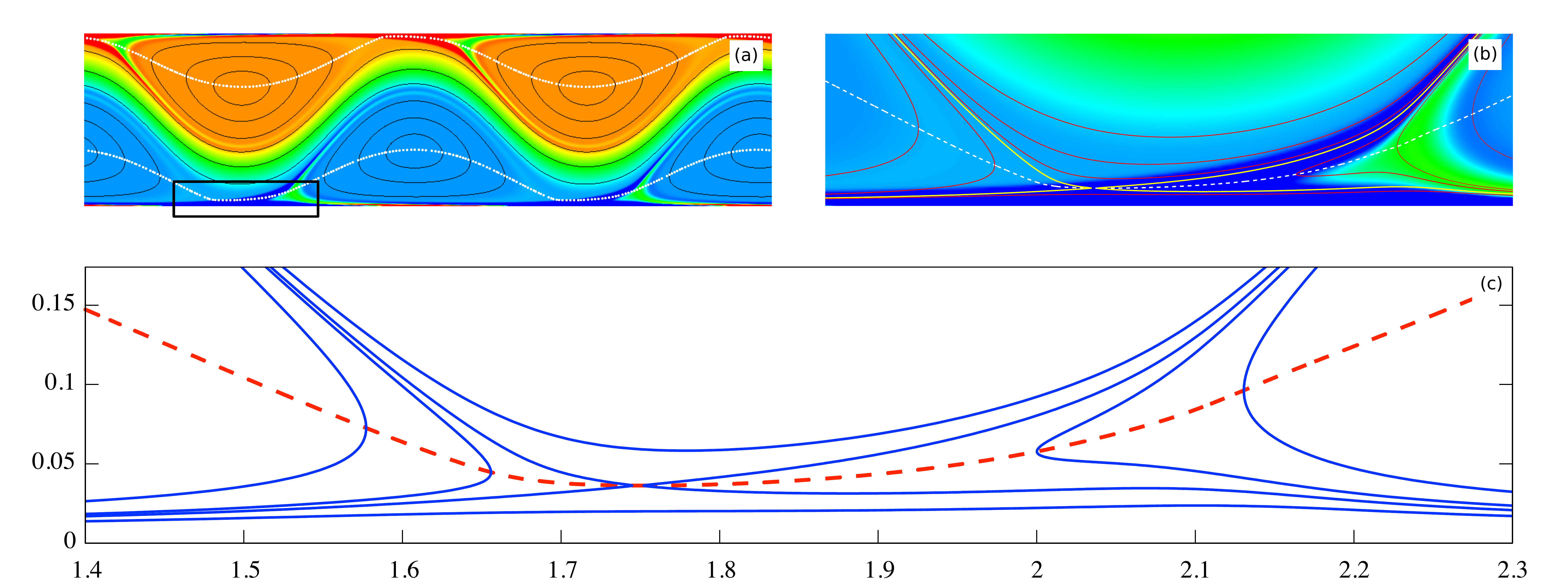}
  \caption{
    (a) Streamlines (solid) and the critical line (broken) for $Re_A=894$ superimposed on the
    vorticity field with box indicating zoom region shown in (b);
    (c) Streamlines and the critical line near the stagnation point
     for $Re_A=516$. }
    \label{stagnation}
\end{figure*}

A natural way to treat appearance of a traveling wave in a parallel
flow is to consider a critical layer where the speeds of the flow and
the wave coincide. For sufficiently small viscosity and friction, each
critical line in 2d is expected to generate a chain of cat-eye
vortices of the same sign.  As one can see in Figure~\ref{jet}, at
small bottom friction, the snake is born with already a finite
amplitude of modulation $H$ and cannot be described as a small
perturbation of a laminar flow, in spite of a sinusoidal shape of the
jet.  Moreover, the whole flow is far from being trivial since the
boundary layer is quite complicated.  The critical line (where the
horizontal velocity is equal to the speed of the traveling wave), of
course, passes through the vortex centers in the traveling-wave
co-moving frame. Yet in between it almost touches the wall, passing
through the stagnation point where two separatrices cross: the one
separating the vortex from the wall and another from the jet, see
Figure~\ref{stagnation}.  Huge velocity gradients are created there,
and a very narrow tongue of negative high vorticity separates from the
wall and goes along the separatrix up into the bulk.  In addition, the
stream of high positive vorticity is generated where the vortex rubs
against the wall in the direction opposite to that of the jet.  On the
way up the streams of positive and negative vorticity blend together
approximately during one vortex turnover time. A significant future
numerical and analytical work will be required to understand and
describe this elaborate structure of the traveling-wave flow and its
evolution with $Re$. (Note that at the coherent sinuous regime, the
jet amplitude $H$ grows, while the width $h$ decreases with the
Reynolds number.)

%\begin{figure}
%  \includegraphics[width=\plotwidth]{stp_pg1e5_img2.pdf}
%  \includegraphics[width=\plotwidth]{stp_pg1e5.pdf}
%  \includegraphics[width=\plotwidth]{stp_pg3e5.pdf}
%  \caption{\addGF{Select one or two out of four panels:}
%    (a)-(b) Streamlines and critical line for $Re_A=894$ superimposed on
%    vorticity field with zoom to the stagnation point.
%    (c)-(d)   Streamlines and critical line near stagnation point
%     for $Re_A=894$ and $Re_A=516$. \addGF{See also a version on the last page.}
%     Distance between from the minimum of critical line moves closer to the wall
%     as $Re_A$ increases, as shown in Fig.~6c.
%     }
%    \label{stagnation}
%\end{figure}

\begin{figure}
  \includegraphics[width=0.5\plotwidth]{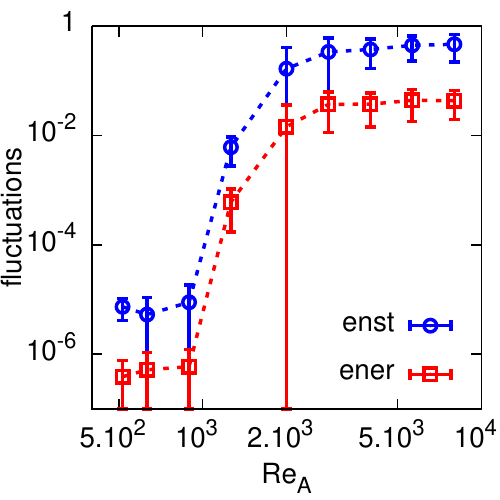}
  \caption{Relative fluctuations in total energy and enstrophy
    in the reference frame of a moving vortex.}
    \label{fluc}
\end{figure}

\begin{figure}
  \includegraphics[width=\plotwidth]{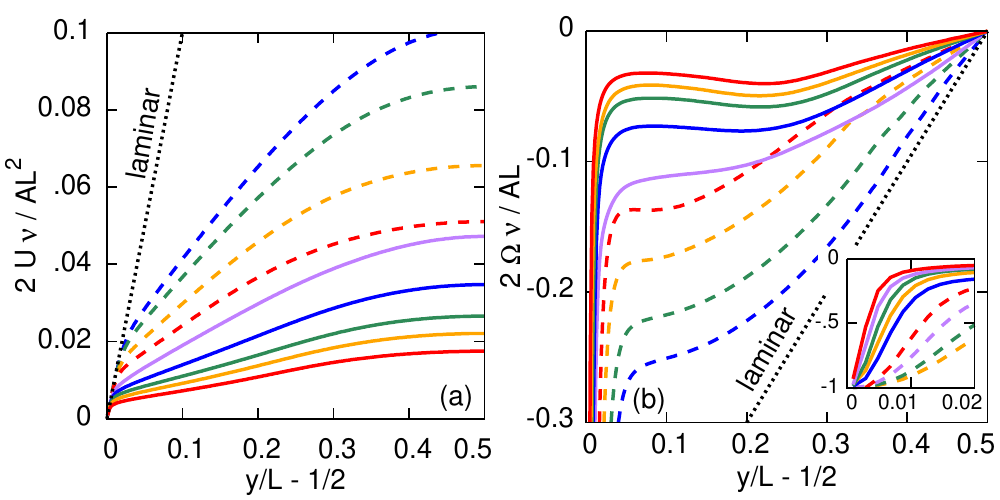}
  \caption{The time averaged and horizontally averaged profiles for
    (a) velocity for $Re_A = 516$, 632, 894, 1264, 2000, 2828, 4000,
    5656, and 8000 (top to bottom) and (b) vorticity for same $Re_A$
    (bottom to top). The velocity is rescaled by the laminar velocity at the
    center and the vorticity is rescaled by the laminar vorticity at the the wall.
    Sinuous and turbulent solutions are shown with dashed and solid
    lines respectively.}
    \label{avg_profiles}
\end{figure}

The description of the sinuous and turbulent flows are presented in the
main text, as well as the discussion of mean flows and fluctuations. The
relative values for fluctuations are shown here in Fig.~\ref{fluc}.
We also supplement the main text with the profiles of the horizontally averaged
velocity and vorticity, shown in Fig.~\ref{avg_profiles}.  The
profiles for the two types of flow are of qualitatively the same
shape: velocity near the wall has the same slope as the laminar profile
for corresponding $Re_A$ and vorticity has the same value.  Away from
the walls, the quantities, normalized to their laminar values,
diminish with $Re_A$.  At the wall, $\Omega$ is very close to the
value given by the laminar parabolic profile.  Surprisingly, the correction
is quadratic in distance, not linear as it would be in classical Poiseuille
flow. % Overall, the transition from low-fluctuating to turbulent
%regime is noticeable, at most, in the appearance of an inflection
%point in vorticity profiles, which is very minor distinction.  To see
%the qualitative difference between sinuous and turbulent flows one must
%look at the distribution of average vorticity in the domain, as we do
%in the main text.

%--------------------------------------

\subsection*{Ball-bearing model}
\label{ball-bearing}

The remarkably simple structure of the flow as a jet slithering
between vortices suggests to treat it with a simple mechanical image
of a ball-bearing.  The jet, like a string pulled between rolls,
rotates the rolls, which like balls in a ball-bearing ``lubricate'' the
system. That allows one to go beyond the scaling and estimate the
order-unity numerical pre-factor.

In the main text we have pointed out that the traveling wave observed at
moderate $Re$ consists of the jet and vortex cells, as illustrated in
Figure~\ref{jet}.  Vorticity is uniform across each cell, $\omega =
\pm \omega_0$, and is linearly distributed inside the jet, $-\omega_0
< \omega < \omega_0$, as shown in Fig.~1a and in
Fig.~4 of the Letter.  We neglect narrow boundary layer near the wall and
assume that circulation around the vortex cell, in the reference frame
of traveling wave, is $s L \simeq g \omega_0 L^2$, where $s$ is the
speed of the traveling wave and $g$ is some geometrical factor.
Inside the jet, we can assume Poiseuille flow, $v(y') = s + ( A/ 2\nu)
y'(h-y')$, where $y'$ is the coordinate perpendicular to the jet, and
$h$ is the thickness of the jet.  Vorticity is continuous at the
border between the vortex and the jet, $\pm \omega_0 \simeq Ah/2\nu
\simeq s/gL$. Thus, for the Reynolds number based on the traveling wave
speed, we obtain,
\[
Re_s \simeq \frac{g}{2} \frac{h}{L} Re^2_A.
\]

This relation connects the speed of the traveling wave with the
applied pressure through the width of the jet. Let us take a closer
look at the geometry of the jet and the length scales of the flow in
our numerical solutions.  It turns out that the level sets of constant
vorticity for $\omega \lesssim \omega_0$ are well described by $y = H
\sin(\pi x/L) + const$, as shown in Fig.~\ref{jet}.  We evaluate the
amplitude of the jet modulation $H$ and the jet width $h$ considering the
level set $\omega = 0$, and the double distance between level sets
$\omega = \pm \omega_0/2$ at the inflection point respectively.  The
width of the boundary layer can be estimated from the average
vorticity profile as $\ell = \sqrt{\Omega/\Omega_{yy}}$.  As shown in
Fig.~\ref{jet}(b,c) the boundary layer is an order of magnitude
narrower than the jet, which justifies our assumption that velocity at
the edge of boundary layer is approximately $s$.  As $Re_A$ increases,
up to onset of fluctuations, the amplitude of the jet increases and
its thickness decreases (which agrees with $2H+h \lesssim
L$).  The decrease of curvature at low $Re$  is consistent
with our observations that such flows favor longer wavelengths.  Until
the onset of fluctuations, the width of the jet scales as $h/L = 17
Re_A^{-2/3} $.

Going back to the ball-bearing model, the scaling for $h/L$ leads to
$Re_s \sim Re_A^{4/3}$. For the proportionality coefficient to be 1.8
(the constant for the blue line in the inset in Fig.~5a in the
Letter) the geometrical factor needs to be $g = 0.21$.  Note that if
the region of constant vorticity were a circle of radius $L/2$, it
would make $g$ equal to 1/4.


\begin{thebibliography}{99}
\bibitem{Rev} B. Eckhardt, T. Schneider, B. Hof, J. Westerweel,
%Turbulence   Transition in Pipe Flow,
{\it Ann  Rev  Fluid Mech} 2007
  39:1, 447-468
\bibitem{Lin}  C. C. Lin, {\it Proc. Nat. Acad. Sci. U.S.A.} {\bf30}, 316-324 (1944)
\bibitem{OK}S. Orszag and L. Kells, %Transition to turbulence in plane
%  Poiseuille and plane Couette flow,
  {\it J  Fluid Mech.}
  {\bf96}, 159-205 (1980)
\bibitem{FE} H. Faisst, B. Eckhardt, %Traveling Waves
%  in Pipe Flow,
  {\it Phys. Rev. Lett.} {\bf91}, 224502 (2003)
\bibitem{WK} H. Wedin and R. Kerswell,  %Exact coherent
%  structures in pipe flow: Travelling wave solutions.
  {\it J  Fluid
  Mech}, {\bf508}, 333-371 (2004). %doi:10.1017/S0022112004009346
\bibitem{BH1} B. Hof et al, %, Casimir W. H. van Doorne1, Jerry Westerweel,
%  Frans T. M. Nieuwstadt, Holger Faisst, Bruno Eckhardt, Hakan Wedin,
%  Richard R. Kerswell, Fabian Waleffe, %Experimental Observation of
%  Nonlinear Traveling Waves in Turbulent Pipe Flow,
  {\it Science} {\bf 305},
  5690, 1594-1598 (2011)%DOI: 10.1126/science.1100393
\bibitem{HH} R. Hewitt and P. Hall,
%The evolution of finite-amplitude wavetrains in plane channel flow
{\it Phil. Trans. R. Soc. Lond. A }  {\bf356}, 2413–2446 (1998)
\bibitem{soap1} Y. Couder, J.M. Chomaz, M. Rabaud, %On the hydrodynamics of soap films
{\it Physica D} {\bf37}, 384 (1989)
\bibitem{soap2} H. Kellay, X-l. Wu and W. Goldburg,% Experiments with Turbulent Soap Films
{\it Phys. Rev. Lett.} {\bf 74}, 3975 (1995)
\bibitem{PhysFluids} %G. Falkovich, G. Boffetta, M. Shats, and A. S. Lanotte, Introduction to
Focus Issue: Two-Dimensional Turbulence, {\it Physics of Fluids} {\bf29}, 110901 (2017)
\bibitem{Pinaki} C. Liu, R. Cerbus and P. Chakraborty, % Janus Spectra in Two-Dimensional Flows
{\it Phys. Rev. Lett.} {\bf117}, 114502 (2016)
\bibitem{Rot} J. P. Rothstein, %Slip on superhydrophobic surfaces-
{\it Ann Rev Fluid Mech} {\bf42}, 89 (2010)
\bibitem{Falkovich} G. Falkovich, {\it Fluid Mechanics, second edition} (Cambridge Univ Press 2018)
\bibitem{Shats} H. Xia, D. Byrne, G. Falkovich, M. Shats, %Upscale energy transfer in thick turbulent fluid layers.
{\it Nature Physics} {\bf7}, 321-324 (2011)
\bibitem{Jimenez} J. Jimenez, %Transition to turbulence in
%  two-dimensional Poiseuille flow,
  {\it J Fluid Mech}, {\bf 218},
 265-297 (1990)
\bibitem{Bat} G. K. Batchelor, %On Steady Laminar Flow with Closed Streamlines at Large Reynolds Number,
{\it J Fluid Mech}  {\bf1,2}, 177-190 (1956)
\bibitem{FR}E Fried, I.E. Idelchik, {\it Flow resistance : a design guide for engineers}, (Hemisphere Pub.  New York  1989)
\bibitem{Childress} S. Childress, %Alpha-effect in flux ropes and
%  sheets,
  {\it Phys. Earth Planet Internat}, {\bf20}, 172-180 (1979).
\bibitem{BS} B. Shraiman, %Diffusive transport in a Rayleigh-Benard
%  convection cell,
  {\it Phys. Rev. A}, {\bf36}, 261 (1987).
  \bibitem{2/3} M. N. Rosenbluth, H. L. Berk, I. Doxas and W. Horton,
  %Effective diffusion in laminar convective flows,
  {\it Phys. Fluids}, {\bf30}
  (1987), pp. 2636-2647.
\bibitem{NV} N. Vladimirova, P. Constantin, A. Kiselev, O. Ruchayskiy
  and L. Ryzhik, %Flame enhancement and quenching in fluid flows
  {\it Combustion Theory and Modelling}, {\bf7}:3, 487-508 (2003)
\bibitem{KoT}K. S. Turitsyn, %Polymer dynamics in chaotic flows with a
%  strong shear component,
{\it J Exp Theor Phys}, {\bf105}, 655-664 (2007);
  arxiv nlin/0501025
\bibitem{Kurchan} K. Lam and J. Kurchan, {\it J Stat Phys} {\bf156}, 619 (2014).



\end{thebibliography}
\end{document}